\documentclass[aps,prb,twocolumn,superscriptaddress]{revtex4-2}
\usepackage{graphicx}
\usepackage{latexsym}
\usepackage{amssymb}
\usepackage{amsmath}
\usepackage{amsfonts}
\usepackage{wasysym}
\usepackage[vcentermath]{youngtab}
\usepackage{upgreek}
\usepackage{bm,dsfont}
\usepackage{multirow}
\usepackage{enumitem}
\usepackage{makecell}
\usepackage{framed}
\usepackage{color}
\usepackage{hyperref}
\hypersetup{
colorlinks = true,
linkcolor = [rgb]{0.70,0.13,0.13},
citecolor = [rgb]{0.13,0.55,0.13},
urlcolor = [rgb]{0.25, 0.41, 0.88}}

\newcommand{\ket}[1]{|#1 \rangle}

\newcommand{\dd}{\mathrm{d}}

\newcommand{\ii}{\mathrm{i}}
\newcommand{\e}{\mathrm{e}}
\newcommand{\tw}{\mathrm{tw}}

\newcommand{\id}{\mathds{1}}

\newcommand{\LU}{\mathrm{LU}}
\newcommand{\U}{\mathrm{U}}
\newcommand{\SU}{\mathrm{SU}}

\newcommand{\Spin}{\mathrm{Spin}}

\newcommand{\dsN}{\mathbb{N}}

\newcommand{\dsR}{\mathbb{R}}

\newcommand{\dsZ}{\mathbb{Z}}

\newcommand{\scK}{\mathcal{K}}

\newcommand{\scO}{\mathcal{O}}

\newcommand{\scT}{\mathcal{T}}

\newcommand{\Tr}{\operatorname{Tr}}
\newcommand{\sgn}{\operatorname{sgn}}
\newcommand{\vol}{\operatorname{vol}}

\newcommand{\vect}[1]{{\bm{#1}}}

\newcommand{\eq}[1]{\begin{equation}#1\end{equation}}
\newcommand{\eqs}[1]{\begin{equation}\begin{split}#1\end{split}\end{equation}}
\newcommand{\eqnref}[1]{Eq.\,\eqref{#1}}
\newcommand{\figref}[1]{Fig.\,\ref{#1}}
\newcommand{\tabref}[1]{Tab.\,\ref{#1}}
\newcommand{\secref}[1]{Sec.\,\ref{#1}}
\newcommand{\refcite}[1]{Ref.\,\onlinecite{#1}}

\newcommand{\bea}{\begin{eqnarray}}
\newcommand{\eea}{\end{eqnarray}}
\def\be{\begin{equation}}
\def\ee{\end{equation}}
\newcommand{\beq}{\begin{equation}}
\newcommand{\eeq}{\end{equation}}
\newcommand{\beqn}{\begin{eqnarray}}
\newcommand{\eeqn}{\end{eqnarray}}

\begin{document}

\title{Definition and Classification of Fermi Surface Anomalies}

\author{Da-Chuan Lu}
\affiliation{Department of Physics, University of California, San Diego, CA 92093, USA}
\author{Juven Wang}
\affiliation{Center of Mathematical Sciences and Applications, Harvard University, Cambridge, {MA 02138}, USA}
\author{Yi-Zhuang You}
\affiliation{Department of Physics, University of California, San Diego, CA 92093, USA}

\date{\today}
\begin{abstract}
We propose that the Fermi surface anomaly of symmetry group $G$ in any dimension is universally classified by $G$-symmetric interacting fermionic symmetry-protected topological (SPT) phases in $(0+1)$-dimensional spacetime. The argument is based on the perspective that the gapless fermions on the Fermi surface can be viewed as the topological boundary modes of Chern insulators in the phase space (position-momentum space). Given the non-commutative nature of the phase space coordinates, we show that the momentum space dimensions should be counted as negative dimensions for SPT classification purposes. Therefore, the classification of phase-space Chern insulators (or, more generally fermionic SPT phases) always reduces to a $(0+1)$-dimensional problem, which can then be answered by the cobordism approach. 
In addition to the codimension-1 Fermi surface case, we also discuss the codimension-$p$ Fermi surface case briefly. We provide concrete examples to demonstrate the validity of our classification scheme, and make connections to the recent development of Fermi surface symmetric mass generation.
\end{abstract}
\maketitle

\tableofcontents

\section{Introduction}

The Fermi liquid \cite{lifshitz2013statistical, pines2018theory} is a conventional and ubiquitous phase of matter in condensed matter physics, modeling the universal low-energy features of electrons in metals. Despite its long history of study, there has been renewed interest in the Fermi liquid, motivated by the quest to understand the surprising stability \cite{Pomeranchuk1958, Gholizade2012, Watanabe1404.3728} of gapless fermions on the Fermi surface. An emerging paradigm in condensed matter theory is to understand all gapless quantum phases of matter from the perspective of emergent symmetries and quantum anomalies \cite{Moon1503.05199, Wang1703.02426, Wen1812.02517, Chen1903.12334, Delacretaz1908.06977, Ji1912.13492, Yang2203.15791, McGreevy2204.03045, Chatterjee2205.06244, Wen2208.09001}. This paradigm has led to significant progress in understanding the Fermi liquid as a gapless state of fermions protected by an emergent quantum anomaly known as the \emph{Fermi surface anomaly} \cite{Watanabe1505.04193, Cheng1511.02263, Lu1705.09298, Cho1705.03892, Bultinck1808.00324, Song1909.08637, Else2007.07896, Else2010.10523, Wen2101.08772, Ma2110.09492, Wang2110.10692, Darius-Shi2204.07585, Lu2210.16304, Cheng2211.12543}.

The boundary-bulk correspondence between quantum anomalies and symmetry-protected topological (SPT) orders has been a key area of study in condensed matter physics in the past decade \cite{Ryu1010.0936,  Wen1303.1803, Kapustin1403.0617, Hsieh1403.6902, Kapustin1404.3230, Wang1405.7689, Else1409.5436,  Hsieh1503.01411, Witten1508.04715, Tiwari1710.04730}. There is a growing consensus \cite{Horava2005FSTopo, Zhao2013FSTopo, Shinsei2013FSTopo, Bulmash1410.4202, Zhang2017FSTopo} 
that the gapless fermions on the Fermi surface can be viewed as the topological boundary modes of a bulk fermionic SPT state and that the Fermi surface anomaly is related to the bulk SPT order. So what on earth should be the ``bulk'' of a Fermi surface? The most honest answer is the Fermi sea --- a region in the momentum space enclosed by the Fermi surface. Then what is ``topological'' about the Fermi sea? \refcite{Bulmash1410.4202} made a key observation that a $d$-dimensional Fermi sea could be viewed as a quantum Hall insulator (or, equivalently, a Chern insulator) in the $2d$-dimensional \emph{phase space} (i.e.~position-momentum space). This sets the basis for classifying Fermi surface anomaly by classifying topological insulators in the phase space.

The main goal of this work is to provide a comprehensive and rigorous classification of the Fermi surface anomaly along the above line of thought. We will primarily consider codimension-1 Fermi surface \cite{Ma2110.09492} (i.e., the Fermi surface is one dimension less than the momentum space dimension) and comment on the higher codimension cases in the summary section (\secref{sec: summary}). Our key result is that the classification of the Fermi surface anomaly in any spacetime dimension is universally equivalent to the classification of interacting fermionic SPT phases in (0+1)-dimensional spacetime. This might not be too surprising as many thermodynamic and transport properties of Fermi liquids remain identical across different dimensions already. The proposed equivalence is established through a careful analysis of the non-commutative geometry \cite{Seiberghep-th/9908142, Dong2006.01282, Connes2014} in phase space, the synthetic dimension reduction \cite{Teo1006.0690, Jian1804.03658} of a phase-space Dirac fermion field theory, and the use of cobordism classification \cite{Kapustin1403.1467, Kapustin1404.6659, Kapustin1406.7329, Freed1604.06527, Guo1711.11587, Wan1812.11967, Yonekura1803.10796, Witten1909.08775, Wan1912.13504, Guo1812.11959} for interacting fermionic SPT states.

We also provide a non-perturbative definition \cite{Cheng2211.12543} of the Fermi surface anomaly protected by the internal symmetry $G$ and the translation symmetry. When $G=\U(1)$, our results match known results such as the Luttinger theorem \cite{LuttingerRP1960, Paramekanticond-mat/0406619, Haldanecond-mat/0505529} for conventional Fermi liquids. When the $\U(1)$ symmetry is broken down to $G=\dsZ_4$ (both contain the fermion parity symmetry ${\dsZ_2^F}$ as a subgroup), we discover non-trivial examples of Fermi surface symmetric mass generation (SMG) \cite{Lu2210.16304}, where the Fermi surface can be gapped out by multi-fermion interactions and deformed to a trivial product state without breaking any symmetry. These novel gapping mechanisms may shed light on the understanding of pseudo-gap physics in cuprates \cite{Zhang2001.09159, Zhang2006.01140}. 

The article is organized as follows. In \secref{sec: effective}, we analyze the non-commutative geometry in the phase space to establish a mathematical foundation for defining quantum field theory in the phase space. We propose a phase-space Dirac fermion field theory as the bulk regularization for the Fermi surface and demonstrate that it reproduces the expected phase space Chern-Simons response theory of the Fermi liquid, as well as the Fermi surface gapless modes as topological boundary modes. This sets the stage for our argument. We then provide a non-perturbative definition of the Fermi surface anomaly and connect it to the recently proposed emergent loop group anomaly in \secref{sec: definition}. Using dimension reduction techniques of synthetic dimensions, we prove our key result: the equivalence between Fermi surface anomaly and (0+1)-dimensional fermionic SPT order in \secref{sec: classification}. We use cobordism tools to classify a list of unitary and anti-unitary symmetries and provide physical insights into our classification results. The article concludes with a summary in \secref{sec: summary}.

\section{Effective Descriptions of Fermi Liquids}\label{sec: effective}

\subsection{Non-Commutative Phase Space Geometry}\label{sec: ncg}

Given the spacetime manifold $M_d\times\dsR$ of a $(d+1)$-dimensional physical system (where $M_d$ is the $d$-dimensional spatial manifold  and $\dsR$ is the time axis), for each position $\vect{x}=(x_1,x_2,\cdots,x_d)\in M_d$ in the space, the conjugate momentum $\vect{k}=(k_1,k_2,\cdots,k_d)$ generates infinitesimal translations on the manifold $M_d$ in the vicinity of $\vect{x}$ and hence lives in the $d$-dimensional cotangent space $T_\vect{x}^*M_d$. Thus the phase space is represented by the cotangent bundle $T^*M_d:=\{(\vect{x},\vect{k})|\vect{x}\in M_d,\vect{k}\in T_\vect{x}^*M_d\}$,
equipped with a canonical commutator (setting $\hbar=1$)
\eq{\label{eq: [x,k]}
[x_i,k_i]=\ii\quad(i=1,2,\cdots,d),}
with $\ii$ being the imaginary unit.
Unlike in a classical space where all coordinates commute, the phase space coordinates obey non-trivial commutation relations \eqnref{eq: [x,k]}, which makes the phase space $T^*M_d$ a non-commutative manifold.

There are two strategies to deal with the non-commutative phase space coordinates:
\begin{itemize}
\setlength\itemsep{0pt}
\item[(I)] \emph{Phase-space background Berry curvature}. Treat both $\vect{x}$ and $\vect{k}$ as ordinary commuting coordinates at the price of introducing a uniform background magnetic field (Berry curvature) in each $(x_i,k_i)$-plane, such that any unit-charged particle moving in such a background magnetic field will accumulate the same Berry phase as required by the commutation relation \eqnref{eq: [x,k]}.

\item[(II)] \emph{Canonical quantization}. Represent the position operator $\vect{x}=\ii\partial_{\vect{k}}$ as a gradient operator in the eigenbasis of the momentum operator $\vect{k}$, or vice versa $\vect{k}=-\ii\partial_{\vect{x}}$, such that the commutation relation \eqnref{eq: [x,k]} is satisfied on the operator level as in quantum mechanics.

\end{itemize}
The strategy (I) of phase-space background Berry curvature has been used in many works \cite{Bulmash1410.4202, Else2007.07896, Else2010.10523, Ma2110.09492, Wang2110.10692} to formulate the Fermi liquid as a phase-space quantum Hall insulator. The phase-space Berry curvature is also responsible for the Berry phase term in Wen's effective theory of Fermi liquid \cite{Wen2101.08772}, or the Wess-Zumino-Witten term in the recently proposed nonlinear bosonization of Fermi surfaces by the coadjoint orbit method \cite{Delacretaz2203.05004}. In this work, we will explore more of the strategy (II) of canonical quantization and hope to gain different insights.

For simplicity, we will always restrict our scope to a \emph{translation invariant} Fermi liquid in the Euclidean position space $M_d=\dsR^d$, then the momentum space is also Euclidean $T_\vect{x}^*M_d=\dsR^d$ and is identical among all points $\vect{x}$. The phase space reduces to a trivial bundle as a product of the position and the momentum spaces
\eq{T^*M_d = \dsR^d\Bowtie\dsR^d.} 
We use the symbol $\Bowtie$ instead of $\times$ to indicate the non-commutative nature between the position and momentum space coordinates.

\subsection{Bulk Description: Fermi Sea = Phase-Space Chern Insulator}\label{sec: blk}

A Chern insulator in the phase space $T^*M_d$ can be formally described by a low-energy effective Hamiltonian of massive Dirac fermions \cite{Bulmash1410.4202}
\eq{\label{eq: H Dirac}
H=\int_{T^*M_d}\dd^d\vect{x}\dd^d\vect{k}\;\psi^\dagger(\ii\partial_{\vect{x}}\cdot\vect{\Gamma}_x+\ii\partial_{\vect{k}}\cdot\vect{\Gamma}_k+m(\vect{k})\Gamma^0)\psi,}
where $\psi:=\psi(\vect{x},\vect{k})$ is a $2^d$-component fermion operator defined at each ``point'' of the 
$2d$-dimensional
phase space $T^*M_d = \dsR^d\Bowtie\dsR^d$ (let us not worry about the non-commutativity between $\vect{x}$ and $\vect{k}$ for now, which will be resolved later). Let $\Gamma^\mu$ (for $\mu=0,1,2,\cdots, 2d$) be a set of $2^d\times 2^d$ anti-commuting Hermitian matrices, satisfying $\{\Gamma^\mu,\Gamma^\nu\}=2\delta^{\mu\nu}$ and $\Gamma^0=\ii^d\prod_{\mu=1}^{2d}\Gamma^\mu$. These $\Gamma$ matrices can be grouped into the temporal $\Gamma^0$, the position spatial $\vect{\Gamma}_x=(\Gamma^1,\cdots,\Gamma^d)$, and the momentum spatial $\vect{\Gamma}_k=(\Gamma^{d+1},\cdots,\Gamma^{2d})$ components. Here $\ii\partial_\vect{x}\cdot\vect{\Gamma}_x=\sum_{i=1}^{d}\ii\partial_{x_i}\Gamma^i$ denotes the dot product between the differential operator $\ii\partial_\vect{x}$ and the set of matrices $\vect{\Gamma}_x$, and similarly for $\ii\partial_\vect{k}\cdot\vect{\Gamma}_k$. A few comments on this theory are as follows:

\begin{itemize}
\setlength\itemsep{0pt}
\item \emph{Locality}. Without interaction, \eqnref{eq: H Dirac} looks like a valid local theory of the fermion field $\psi$ in the phase space. However, once fermion interaction is introduced, \eqnref{eq: H Dirac} is no longer a local field theory because the interaction is generally non-local in the momentum space. Therefore, \eqnref{eq: H Dirac} should only be viewed as a ``formal'' description of the phase-space Chern insulator. One way to regularize the theory is to evoke the strategy (II) in \secref{sec: ncg} to resolve the non-commutative phase space geometry by replacing $\ii\partial_\vect{k}\to\vect{x}$, and rewrite \eqnref{eq: H Dirac} as
\eq{\label{eq: H Dirac reg}
H=\int_{M_d}\dd^d\vect{x}\;\psi^\dagger(\ii\partial_{\vect{x}}\cdot\vect{\Gamma}_x+\vect{x}\cdot\vect{\Gamma}_k+m(-\ii\partial_\vect{x})\Gamma^0)\psi,}
which is solely defined in the position space and respects the position space locality such that local interactions can be introduced if needed. 

\item \emph{Mass profile}. The bulk Dirac mass $m(\vect{k})$ is supposed to be a polynomial function of $\vect{k}$, which specifies the shape of the Chern insulator in the phase space. For example, given the Fermi momentum $k_F$, $m(\vect{k})=\vect{k}^2-k_F^2$ is one possible choice of the mass profile. Suppose the Fermi sea occupies a region $\Omega\subset \dsR^d$ in the momentum space enclosed by the $(d-1)$-dimensional Fermi surface $\partial \Omega$, the Dirac fermion mass profile should satisfy
\eq{\label{eq: m profile}
m(\vect{k})\left\{
\begin{array}{cc}
\leq 0 & \text{if }\vect{k}\in \Omega,\\
>0 & \text{if }\vect{k}\notin \Omega.
\end{array}\right.}
This described a phase-space Chern insulator in the Fermi sea region $\Omega$, such that the Fermi surface $\partial\Omega$ (as the boundary of the phase-space Chern insulator) corresponds to the mass domain wall at $m(\vect{k})=0$. 

The fermions are gapped everywhere in the phase space except on the Fermi surface, where the fermion mass vanishes. This is consistent with the physical intuition that the gapless fermions on the Fermi surface are the only non-trivial low-energy feature of the Fermi liquid. We will study these boundary fermion modes in more detail in \secref{sec:bdy} to show that they travel in the directions perpendicular to the Fermi surface as expected.

\item \emph{Particle-hole symmetry}. Under the particle-hole transformation $\dsZ_2^C$, the inside and outside of the Fermi surface will interchange, corresponding to flipping the fermion mass $\dsZ_2^C: m\to -m$, or equivalently, conjugating the fermion operator 
\eq{\dsZ_2^C: \psi\to \scK \Gamma^0\psi^*,}
where $\scK$ denotes the complex conjugate operator, such that $\dsZ_2^C: \psi^\dagger\Gamma^0\psi\to -\psi^\dagger\Gamma^0\psi$. Note that $\dsZ_2^C$ is \emph{not} a symmetry of the Hamiltonian $H$ in \eqnref{eq: H Dirac}, as the mass term $m$ explicitly breaks this symmetry. However, it is useful in defining the Fermi surface. We propose that the Fermi surface should be more generally defined as the \emph{particle-hole symmetric} sub-manifold in the phase space, specified by the locus of $\langle \psi^\dagger \Gamma^0\psi \rangle =0$. This definition applies to the case of interacting fermions.

\item \emph{Phase-space $\U(1)$ symmetry}. The Hamiltonian $H$ in \eqnref{eq: H Dirac} has a 0-form $\U(1)$ symmetry in the phase space, generated by the charge operator
\eq{\label{eq: Q}
Q=\int_{T^*M_d}\dd^d\vect{x}\dd^d\vect{k}\;\psi^\dagger \psi.}
The symmetry transformation $\e^{\ii \phi Q}$ forms the $\U(1)$ symmetry group,
where $\phi \in [0, 2 \pi)$ and $Q \in \mathbb{Z}$.
The fermion field transforms as $\psi\to\e^{\ii\phi}\psi$ under the symmetry transformation.
\end{itemize}

The essential bulk topological response of the Fermi liquid is captured by a phase-space Chern-Simons theory \cite{Bulmash1410.4202, Else2007.07896, Else2010.10523, Ma2110.09492, Wang2110.10692} of the phase-space $\U(1)$ symmetry. To show that the effective Hamiltonian in \eqnref{eq: H Dirac} indeed reproduces the desired topological response, we first gauge the 0-form $\U(1)$ symmetry of the fermion $\psi$ (under which $\psi\to\e^{\ii\phi}\psi$) by introducing a 1-form gauge field $A$ in the phase spacetime
\eq{A=A_0\dd t + \vect{A}_x\cdot\dd \vect{x}+\vect{A}_k\cdot\dd \vect{k},}
where $A_0$, $\vect{A}_x=(A_1,\cdots,A_d)$, $\vect{A}_k=(A_{d+1},\cdots,A_{2d})$ are respectively the components of the $\U(1)$ gauge connection in the time, position, and momentum spaces. We will treat $A$ as a background gauge field that does not have dynamics. Let $F:=\dd A$ be the $\U(1)$ gauge curvature. Following the strategy (I) mentioned in \secref{sec: ncg}, we must set $F_{i,d+i}=1$ for $i=1,2,\cdots,d$ to reproduce the position-momentum commutator in \eqnref{eq: [x,k]}. This background gauge curvature effectively replaces the non-commutative $2d$-dimensional
phase space geometry, and the effective Hamiltonian \eqnref{eq: H Dirac} becomes \cite{Bulmash1410.4202}
\eq{\label{eq: H Dirac gauged}
H=\int_{T^*M_d}\dd^d\vect{x}\dd^d\vect{k}\;\psi^\dagger(\ii D_\vect{x}\cdot\vect{\Gamma}_x+\ii D_\vect{k}\cdot\vect{\Gamma}_k+m\Gamma^0-A_0)\psi,}
where $\ii D_\mu:=\ii \partial_\mu-A_\mu$ are gauge covariant derivatives. Now, in \eqnref{eq: H Dirac gauged}, $\vect{x}$ and $\vect{k}$ are ordinary \emph{commuting} coordinates, as the background Berry curvature $F_{i,d+i}=1$ has been implemented in the $\U(1)$ gauge configuration to resolve the non-commutativity. Therefore, we can use conventional field theory approaches to deal with \eqnref{eq: H Dirac gauged}.

Integrating out the fermion field in \eqnref{eq: H Dirac gauged} generates the following Chern-Simons action in the $(2d+1)$-dimensional
phase spacetime \cite{Bulmash1410.4202, Hayata1701.04012} (assuming the Dirac fermion $\psi$ is such regularized that $m>0$ corresponds to a trivial insulator)
\eq{\label{eq: CS}
S=\frac{1}{(d+1)!(2\pi)^d}\int_{T^*M_d\times\dsR}\frac{1-\sgn m}{2}A\wedge(\dd A)^{\wedge d}.}
This is the defining bulk topological field theory whose inflow generates the Fermi surface anomaly \cite{Else2007.07896, Else2010.10523, Darius-Shi2204.07585}. \footnote{Our discussion here is unrelated to the previous study of chiral and gravitational anomalies on Fermi surfaces \cite{Basar1307.2234}, which is about the non-trivial Berry curvature on Fermi surfaces purely defined in the momentum space.} In particular, if we plug in the phase-space background gauge configuration $F_{i,d+i}=1$ (i.e., $\dd A=F=\sum_{i=1}^d\dd x_i\wedge \dd k_i$), take the fermion mass profile in \eqnref{eq: m profile}, and finish the momentum space integration, \eqnref{eq: CS} will reduce to
\eq{S=\frac{1}{(2\pi)^d}\int_{M_d\times\dsR}\dd t\,\dd^d\vect{x}\;A_0 \vol \Omega,}
which indicates that the fermion charge density $\nu$ (filling fraction) is related to the Fermi volume $\vol\Omega$ by
\eq{\nu=\frac{\delta S}{\delta A_0}=\frac{\vol\Omega}{(2\pi)^d},}
where $\vol\Omega:=\int\dd^d\vect{k}\;(1-\sgn m(\vect{k}))/2$ is by-definition the momentum-space volume where $m(\vect{k})\leq 0$.
This is precisely the Luttinger theorem --- a hallmark of the Fermi surface anomaly. Thus we have confirmed that the effective bulk Hamiltonian \eqnref{eq: H Dirac} can produce the correct anomaly inflow to describe a $(d+1)$-dimensional unit-charged Fermi liquid with a single Fermi surface. The extension to cases of generic fermion charges and multiple Fermi surfaces is straightforward (see \refcite{Else2007.07896} for example) and will not be elaborated further here.

\subsection{Boundary Description: Fermi Surface = Phase-Space Chiral Boundary Fermions}\label{sec:bdy}

How do we see more explicitly that the effective Hamiltonian \eqnref{eq: H Dirac} reproduces the low-energy fermions on a Fermi surface? Since the Fermi surface is interpreted as the boundary of the phase-space Chern insulator, the gapless fermions should arise as the topological boundary modes, which can be analyzed as follows.

\begin{figure}[htbp]
\begin{center}
\includegraphics[scale=0.65]{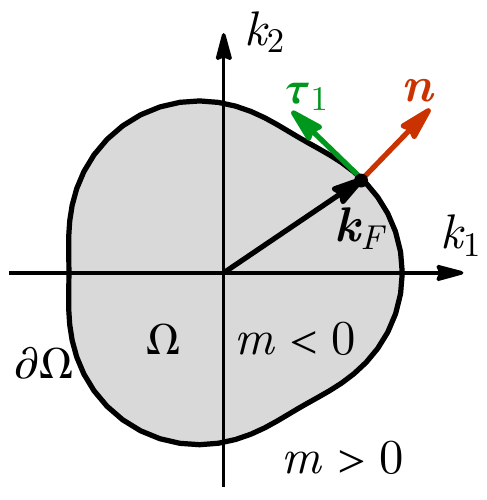}
\caption{Illustration of a point $\vect{k}_F$ on the Fermi surface $\partial\Omega$ with the normal vector $\vect{n}$ and the tangent vector(s) $\vect{\tau}_j$, for the case when the Fermi sea $\Omega$ dimension is $d=2$.}
\label{fig: boundary}
\end{center}
\end{figure}

As shown in \figref{fig: boundary}, we consider a point $\vect{k}_F\in\partial\Omega$ on the Fermi surface at which the normal vector is specified by $\vect{n}$. This means that the fermion mass will cross zero in the phase space at $\vect{k}_F$ with a gradient along the $\vect{n}$ direction:
\eq{\label{eq: m dw}
m(\vect{k}_F)=0,\quad \partial_{\vect{k}}m(\vect{k}_F)\propto \vect{n}.}
Such a mass domain wall at $\vect{k}_F$ will trap gapless fermion modes in the eigenspace specified by the projection $P_0=(1+\ii(\vect{n}\cdot\vect{\Gamma}_k)\Gamma^0)/2$. Under this projection, only those terms that commute with $P_0$ can remain, so the effective Hamiltonian \eqnref{eq: H Dirac} reduces to
\eqs{\label{eq: H bdy1}
H&=\int_{\partial\Omega}\dd\vect{k}_F\int_{M_d\Bowtie T_{\vect{k}_F}\partial\Omega}\dd^d\vect{x}\,\dd^{d-1}\vect{k}\\
&\psi^\dagger P_0\Big(\ii(\vect{n}\cdot\partial_\vect{x})(\vect{n}\cdot\vect{\Gamma}_x)+\\
&\sum_{j=1}^{d-1}\big(\ii(\vect{\tau}_j\cdot\partial_\vect{x})(\vect{\tau}_j\cdot\vect{\Gamma}_x)+\ii(\vect{\tau}_j\cdot\partial_{\vect{k}})(\vect{\tau}_j\cdot\vect{\Gamma}_k)\big)\Big)P_0 \psi,}
where $T_{\vect{k}_F}\partial\Omega$ denotes the $(d-1)$-dimensional tangent space of the Fermi surface $\partial \Omega$ at the base point $\vect{k}_F$, and $\vect{\tau}_j$ (for $j=1,2,\cdots,d-1$) denote a set of orthonomal basis of the tangent space $T_{\vect{k}_F}\partial\Omega$.

To resolve the non-commutativity between $\vect{x}$ and $\vect{k}$ coordinates, we evoke the strategy (II) outlined in \secref{sec: blk}. Given that $\vect{x}=\ii\partial_\vect{k}$ resolves the canonical commutation relation in \eqnref{eq: [x,k]}, we can simply replace the gradient operator $\ii\partial_\vect{k}$ by $\vect{x}$, and fall back to the standard quantum mechanical description in the position space $M$ alone. Under this replacement, \eqnref{eq: H bdy1} becomes
\eqs{\label{eq: H bdy2}
H&=\int_{\partial\Omega}\dd\vect{k}_F\int_{M_d}\dd^d\vect{x}\;\psi^\dagger P_0\Big(\ii(\vect{n}\cdot\partial_\vect{x})(\vect{n}\cdot\vect{\Gamma}_x)\\
&\sum_{j=1}^{d-1}\big(\ii(\vect{\tau}_j\cdot\partial_\vect{x})(\vect{\tau}_j\cdot\vect{\Gamma}_x)+(\vect{\tau}_j\cdot\vect{x})(\vect{\tau}_j\cdot\vect{\Gamma}_k)\big)\Big)P_0 \psi.}
Now the terms $(\vect{\tau}_j\cdot\vect{x})(\vect{\tau}_j\cdot\vect{\Gamma}_k)$ in the Hamiltonian \eqnref{eq: H bdy2} can be interpreted as a new set of perpendicular domain walls of fermion masses (each one is normal to a $\vect{\tau}_j$ direction). They will further localize the fermions to the origin in all tangent directions $\vect{\tau}_j$ (for $j=1,\cdots,d-1$). The localized fermion modes are specified by a sequence of further projections $P_j=(1+(\vect{\tau}_j\cdot\vect{\Gamma}_x)(\vect{\tau}_j\cdot\vect{\Gamma}_k))/2$, such that the total projection is
\eq{
P=P_0\prod_{j=1}^{d-1}P_j.}
Under the total projection $P$, only one fermion mode survives. This can be seen by a simple counting argument: the fermion field $\psi$ has $2^d$ components to start with, given $P_0,\cdots, P_{d-1}$ are $d$ commuting projectors, each reducing the number of fermion components by half, the remaining component number is $2^d/2^d=1$. 

The only term in the Hamiltonian that commute with the total projection $P$ is $\ii(\vect{n}\cdot\partial_\vect{x})(\vect{n}\cdot\vect{\Gamma}_x)$, which will survive in the low-energy theory. Moreover, $(\vect{n}\cdot\vect{\Gamma}_x)$ becomes an identity operator in the projected subspace, because
\eqs{\label{eq: PGP=P}
&P(\vect{n}\cdot\vect{\Gamma}_x)P\\
=&P(\vect{n}\cdot\vect{\Gamma}_x)\ii(\vect{n}\cdot\vect{\Gamma}_k)\Gamma^0\prod_{j=1}^{d-1}\big(\ii(\vect{\tau}_j\cdot\vect{\Gamma}_x)(\vect{\tau}_j\cdot\vect{\Gamma}_k)\big)P\\
=&P\Big(\ii^d\prod_{\mu=0}^{2d}\Gamma^\mu \Big)P=P\id P=P.}
The first equality in \eqnref{eq: PGP=P} relies on the fact that we can insert between projection operators $P$ matrices like $\ii(\vect{n}\cdot\vect{\Gamma}_k)\Gamma^0$ or $\ii(\vect{\tau}_j\cdot\vect{\Gamma}_x)(\vect{\tau}_j\cdot\vect{\Gamma}_k)$, as they all behave like identity operators in the projected subspace. If we denote the projected fermion mode as $\psi_{\vect{k}_F}=P\psi$ (the low-energy fermion localized on the intersection of mass domain walls at the Fermi momentum $\vect{k}_F$), the effective Hamiltonian for this fermion mode reads
\eq{\label{eq: H chiral bdy}
H=\int_{\partial\Omega}\dd\vect{k}_F\int\dd(\vect{n}\cdot\vect{x})\;\psi_{\vect{k}_F}^\dagger \ii(\vect{n}\cdot\partial_\vect{x})\psi_{\vect{k}_F},}
which describes a single chiral fermion moving along the normal direction $\vect{n}$ at every momentum $\vect{k}_F\in\partial\Omega$ on the Fermi surface, which matches the low-energy physics of Fermi liquid precisely. Therefore, the phase-space Chern insulator effective Hamiltonian $H$ in \eqnref{eq: H Dirac} indeed provides a bulk regularization for the Fermi liquid, reproducing all the expected low-energy behaviors of gapless fermions on the Fermi surface. This is an alternative bulk regularization of Fermi liquid compared to the Weyl fermion regularization proposed by Ma and Wang \cite{Ma2110.09492} recently. To make a comparison between our regularization and that in \refcite{Ma2110.09492},
\begin{itemize}
\setlength\itemsep{0pt}
\item We use the canonical quantization approach to resolving the non-commutative phase space geometry, while \refcite{Ma2110.09492} uses the phase-space background Berry curvature approach.
\item The low-energy chiral fermions are realized as domain-wall fermions in our approach, compared to Landau-level Weyl fermions in \refcite{Ma2110.09492}. The directional nature of the chiral fermions (i.e., they always move along the normal direction at each point on the Fermi surface) is more explicit in our regularization.
\end{itemize}

\section{Definition of Fermi Surface Anomaly}\label{sec: definition}

\subsection{Emergent Loop Group Symmetry and Perturbative Fermi Surface Anomaly}\label{sec: loop group}

The chiral boundary fermion effective Hamiltonian \eqnref{eq: H chiral bdy} has a rather large emergent symmetry, described by the loop-$\partial\Omega$ group of $\U(1)$ \cite{Else2007.07896,Else2010.10523} or the mapping space from the Fermi surface $\partial\Omega$ to $\U(1)$, denoted as $\mathrm{L}_{\partial\Omega}\U(1):=\mathrm{Map}(\partial\Omega, \U(1))$ \footnote{For codimension-1 Fermi surface, $\partial\Omega$ is a $(d-1)$-dimensional closed manifold. In the case that $\partial\Omega$ is diffeomorphic to a $S^{d-1}$ sphere, the loop group is also denoted as $\mathrm{L}^{d-1}\U(1)$}. Under the group action, fermion operators transform as
\eq{
\mathrm{L}_{\partial\Omega}\U(1):\psi_{\vect{k}_F}\to\e^{\ii\phi(\vect{k}_F)}\psi_{\vect{k}_F}\quad(\forall\vect{k}_F\in\partial\Omega)}
where $\phi(\vect{k}_F)$ is a continuous function on the Fermi surface $\partial\Omega$, subject to the equivalence $\phi(\vect{k}_F)\sim \phi(\vect{k}_F)+2\pi$. Mathematically, the loop group $\mathrm{L}_{\partial\Omega}\U(1)$ is the group of all continuous maps from the closed manifold $\partial\Omega$ to $\U(1)$, with the group multiplication defined pointwise.

In contrast, for a conventional real-space $\U(1)$-symmetric Chern insulator, the boundary theory only has the same $\U(1)$ symmetry inherited from the bulk. In this case, the boundary symmetry is not enlarged because the gapless fermion mode can propagate (along tangent directions) throughout the boundary, locking point-wise $\U(1)$ transformations together into a global $\U(1)$ transformation on the boundary manifold. However, for the phase-space Chern insulator, due to the non-commutative nature between the position and momentum coordinates, the boundary fermion mode is localized in all tangent directions of the Fermi surface and only propagates along the normal direction $\vect{n}$. Therefore, the $\U(1)$ transformations at different momentum points $\vect{k}_F$ on the Fermi surfaces are not locked together, giving rise to the enlarged loop group symmetry $\mathrm{L}_{\partial\Omega}\U(1)$.

Our argument establishes the loop group symmetry $\mathrm{L}_{\partial\Omega}\U(1)$ on the Fermi surface as an emergent symmetry, originated from the $\U(1)$ symmetry in the phase-space bulk. Therefore, the Fermi surface anomaly, which was proposed \cite{Else2007.07896} to be a perturbative anomaly of $\mathrm{L}_{\partial\Omega}\U(1)$, can be described by the bulk topological field theory of a $\U(1)$ connection $A$ of 
the $\U(1)$ bundle in the phase spacetime, as derived in \eqnref{eq: CS} already,
\eq{\label{eq: CS bulk_restricted}
S=\frac{k}{(d+1)!(2\pi)^d}\int_{M_d\times\Omega\times\dsR}A\wedge(\dd A)^{\wedge d}.}
Here we have added in the Chern-Simons level $k\in\dsZ$ for generality, which should correspond to the multiplicity (degeneracy) of the Fermi surface. We set $k=1$ for a single Fermi surface. Various physical consequences of this theory have been discussed in the literature \cite{Bulmash1410.4202, Else2007.07896, Else2010.10523, Ma2110.09492, Wang2110.10692}, which we will not reiterate. This description sets the basis to classify the loop group $\mathrm{L}G$ anomaly on the $(d-1)$-dimensional Fermi surface by the $G$-symmetric invertible topological phases in the $2d$-dimensional phase space, which will be our key strategy in \secref{sec: classification}.

\subsection{Interstitial Defect and Non-Perturbative Fermi Surface Anomaly}\label{sec: defect}

\begin{figure}[htbp]
\begin{center}
\includegraphics[scale=0.65]{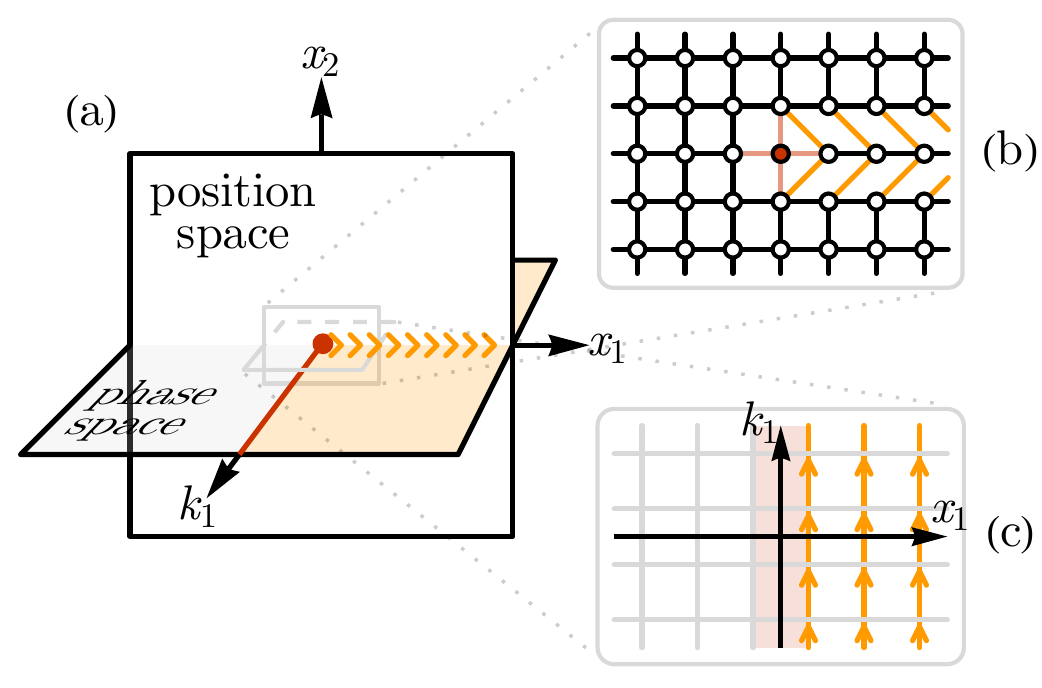}
\caption{(a) Characterize the Fermi surface anomaly by the projective representation of the internal symmetry in the presence of an interstitial defect. (b) On the lattice, an interstitial defect (the red dot) is created by translating a semi-infinite line of sites along the line direction. (c) In the phase space, this creates extra Berry curvature (in the shaded plaquettes) along a line of momenta at the defect position.}
\label{fig: defect}
\end{center}
\end{figure}

One drawback of using the phase-space Chern-Simons theory \eqnref{eq: CS bulk_restricted} to characterize the Fermi surface anomaly is that it is not straightforward to extend the description to Fermi liquids with a more general symmetry group $G$, such as $G=\dsZ_{2n}$. We propose to define the Fermi surface anomaly in a lattice fermion system by the projective representation of the internal symmetry $G$ in the presence of an \emph{interstitial defect} that adds an extra site to the lattice \cite{Cheng1804.10122, Cheng2211.12543}, as illustrated in \figref{fig: defect} (a).

Consider a lattice fermion system in $d$-dimensional space with global internal symmetry $G$ and lattice translation symmetry $\dsZ^d$. Let $T_i$ be the generator of translation symmetry in the $i$-th spatial direction. In the phase space, the lattice translation symmetry $\dsZ^d$ acts as an emanant momentum-space dipole symmetry $\U(1)^d$ (i.e.~the dipole moment conservation in the momentum space)
\eq{\label{eq: dipole sym} \psi(\vect{x},\vect{k})\to T_i\psi(\vect{x},\vect{k}) T_i^{-1}=\e^{\ii k_i}\psi(\vect{x},\vect{k}).}
An emanant symmetry \cite{Cheng2211.12543} is an exact IR symmetry that only acts on low-energy degrees of freedom. Its action on high-energy degrees of freedom is not well-defined. However, it arises from a UV symmetry in that any low-energy operator charged under the emanant symmetry must also be charged under the corresponding UV symmetry. The momentum-space dipole symmetry $\U(1)^d$ in \eqnref{eq: dipole sym} emanates from the lattice translation symmetry $\dsZ^d$ in the sense that any low-energy operator violating the momentum-space dipole symmetry will also break the lattice translation symmetry \cite{Cheng2211.12543, Metlitski1707.07686}, even though there is no group homomorphism between these two symmetry groups. A similar discussion also appeared in \refcite{Wen2101.08772}, where the emanant symmetry was proposed to be $\dsR^d$ (as a non-compact version of our proposed $\U(1)^d$).


An interstitial defect is a point defect that adds one extra site (or unit cell) to the lattice. It can be created by translating a semi-infinite line of sites along the line direction as shown in \figref{fig: defect}(b) on the lattice level. The choice of direction for this semi-infinite line does not matter. We may choose it to be along the positive axis of $x_1$. The twist operator $U_\tw$ creates the interstitial defect at the origin $\vect{x}=0$,
\eq{U_\tw=T_1^{\Theta(x_1)\prod_{i=2}^{d}\delta(x_i)},}
where $\Theta$ is the Heaviside step function and $\delta$ is the Kronecker delta function:
\eq{\Theta(x)=\left\{\begin{array}{ll}1 & \text{if }x>0,\\ 0& \text{if }x<0,\end{array}\right.\quad\delta(x)=\partial_x\Theta(x).}
They together ensure that the translation is only implemented along the positive axis of $x_1$.

For any field or operator $\scO$, we defined the twisted version $\scO_\tw$ as $\scO_\tw:=U_\tw\scO U_\tw^{-1}$. In particular, the fermion field is twisted to
\eq{\label{eq: psi twist}
\psi_\tw(\vect{x},\vect{k})=\e^{\ii k_1\Theta(x_1)\prod_{i=2}^{d}\delta(x_i)}\psi(\vect{x},\vect{k}).}
This allows us to define the twisted Hamiltonian $H_\tw$ and the twisted representation of symmetry operation $\rho_\tw(g)$ for any group element $g\in G$ of the internal symmetry group $G$ by replacing all operators in $H$ or $\rho(g)$ with their twisted version. 
We say that the fermion system has a Fermi surface anomaly, if there exists a cyclic subgroup of $G$ (generated by $g\in G$ and $g^n=1$) such that the twisted partition function accumulates a non-trivial phase $\e^{\ii2\pi\nu}\neq 1$ (or equivalently, a non-trivial index $\nu\neq 0\mod 1$) under the cyclic symmetry action:
\eq{\label{eq: FSA}\Tr (\e^{-\beta H_\tw} \rho_\tw(g)^n)=\e^{\ii2\pi\nu} \Tr \e^{-\beta H_\tw}.}
This indicates that the interstitial defect transforms projectively under the internal symmetry $G$, which provides a non-perturbative definition of the Fermi surface anomaly. From this perspective, the Fermi surface anomaly may also be viewed as the mixed anomaly between the internal symmetry $G$ and the emanant symmetry $\U(1)^d$, which is a straightforward generalization of the mixed $\U(1)\times\dsR^d$ anomaly proposed by Wen \cite{Wen2101.08772}.

To demonstrate the validity of the general definition of the Fermi surface anomaly by \eqnref{eq: FSA}, we consider the special case of $G=\U(1)$ and show that it reproduces the known filling constraints by the Luttinger theorem.
When $G=\U(1)$, for $g=\e^{\ii \phi}\in G$, we have $\rho(g)_\tw=\e^{\ii \phi Q_\tw}$, where $Q_\tw$ is twisted from the charge operator $Q$ in \eqnref{eq: Q}. The twisted partition function can be defined as
\eq{\label{eq: Z_tw U1}
Z_\tw(\beta,\phi)=\Tr (\e^{-\beta H_\tw}\e^{\ii \phi Q_\tw}).}
The Fermi surface anomaly is manifested by
\eq{\label{eq: FSA U1}
Z_\tw(\beta,\phi+2\pi)=\e^{\ii 2 \pi \nu}Z_\tw(\beta,\phi),}
where $\nu$ (mod 1) serves as the anomaly index, and $\e^{\ii 2 \pi \nu}$ is the same non-trivial phase factor that appeared in \eqnref{eq: FSA}. 

To compute the anomaly index $\nu$, we notice that the transformation of the fermion field in \eqnref{eq: psi twist} induces a $\U(1)$ gauge transformation in the phase space \cite{Metlitski1707.07686, Song1909.08637}, such that
\eq{A_\tw=A+\Theta(x_1)\prod_{i=2}^{d}\delta(x_i)\,\dd k_1.}
This means that the background $\U(1)$ gauge field component $A_{k_1}$ in the phase space is shifted by a uniform amount over the half-plane of $x_1>0$, as shown in \figref{fig: defect}(c). As a result, this leads to additional $\U(1)$ gauge curvature $F:=\dd A$ in the phase space along the interface of $x_1=0$,
\eq{F_\tw=F+\delta(\vect{x})\,\dd x_1\wedge \dd k_1,}
where $\delta(\vect{x})=\prod_{i=1}^{d}\delta(x_i)$. Substitute into the bulk topological response theory in \eqnref{eq: CS bulk_restricted}, and take a phase-space uniform configuration for the temporal gauge field $A_0(\vect{x},\vect{k})=\varphi\delta(t)$ at the $t=0$ time slice, we have
\eq{S_\tw=S+k \varphi\frac{\vol\Omega}{(2\pi)^d},}
hence the twisted charge operator is given by
\eq{\label{eq: Q_tw}
Q_\tw=\frac{\partial S_\tw}{\partial \varphi}=Q+k\frac{\vol\Omega}{(2\pi)^d}.}
Substitute \eqnref{eq: Q_tw} to \eqnref{eq: Z_tw U1}, we can compute the anomalous phase factor in \eqnref{eq: FSA U1}. Given that the total charge $Q\in \dsZ$ is quantized, the Fermi surface anomaly index $\nu$ is associated with the fractional charge of the global $\U(1)$ symmetry induced by the interstitial defect
\eq{\nu=k\frac{\vol\Omega}{(2\pi)^d}\mod 1.}
The level $k\in \dsZ$ is integer classified in this case. For generic Fermi volume $\vol\Omega$ that is not a rational fraction of the Brillouin zone volume $(2\pi)^d$, the Fermi surface anomaly is non-vanishing as long as the level $k\neq 0$. This reproduces the know results about Fermi liquid with $\U(1)$ symmetry and demonstrates that our non-perturbative definition of the Fermi surface anomaly in \eqnref{eq: FSA} falls back to the perturbative $\mathrm{L}_{\partial\Omega}\U(1)$ anomaly proposed in \refcite{Else2007.07896, Else2010.10523} for the case of $G=\U(1)$.

For more general internal symmetry $G$, we proposed that the Fermi surface anomaly should be defined via \eqnref{eq: FSA}, following the general idea of the twist defect construction by Cheng and Seiberg \cite{Cheng2211.12543}. The major difference is that they twist the translation symmetry in time and internal symmetry in space, while we twist the translation symmetry in space and internal symmetry in time. This modification allows us to define the Fermi surface anomaly in general dimensions (beyond $(1+1)$D).

\section{Classification of Fermi Surface Anomaly}\label{sec: classification}

\subsection{Synthetic Dimension Reduction Argument}\label{sec: dim}

The remaining objective is to classify the Fermi surface anomaly for a general internal symmetry group $G$. According to \secref{sec: defect}, the anomaly is defined by the fractionalized representation of $G$ carried by interstitial defects in the fermionic system, indicating that the anomaly classification can be mapped to the classification of $(0+1)$-dimensional phase transitions between $G$-symmetric invertible topological phases of fermions, which is equivalent to the classification of $(0+1)$-dimensional fermionic SPT states. However, in \secref{sec: loop group}, the bulk topological field theory described by \eqnref{eq: CS bulk_restricted} suggests a different conclusion that classifying the Fermi surface anomaly of a $(d+1)$-dimensional Fermi liquid is equivalent to classifying the $(d+d+1)$-dimensional fermionic SPT states in phase spacetime. This raises a paradox as the two different counting of dimensions seem to be inconsistent with each other.

The paradox can be resolved by considering the non-trivial dimension counting in the phase space. Because position and momentum are non-commuting coordinates, their dimensions should not be simply added together. Instead, the correct classification should consider the momentum dimensions as ``negative'' spatial dimensions \cite{Teo1006.0690}, effectively defining the bulk SPT phase in a $(d-d+1)=(0+1)$-dimensional spacetime, aligning with the view from the interstitial defect.

To understand this unusual dimension counting, we revisit the effective bulk Hamiltonian in \eqnref{eq: H Dirac}, which describes a phase-space Chern insulator (or, equivalently, a $d$-dimensional Fermi sea). Following the strategy (II) of canonical quantization to regularize the bulk Hamiltonian by replacing $\ii\partial_\vect{k}\to\vect{x}$ as \eqnref{eq: H Dirac reg}, we have
\eq{\label{eq: H synthetic}
H=\int_{M_d}\dd^d\vect{x}\;\psi^\dagger(\ii\partial_{\vect{x}}\cdot\vect{\Gamma}_x+\vect{x}\cdot\vect{\Gamma}_k+m\Gamma^0)\psi.
}
This describes a series of perpendicular mass domain walls (one in each independent direction) that intersect at $\vect{x}=0$, trapping a single fermion mode at the intersection point, which is described by the following effective Hamiltonian: 
\eq{\label{eq: H0d}
H=m(\psi^\dagger\psi-1/2),}
where $m$ plays the role of the chemical potential, and the $2^d$-component spinor is projected to $1$-dimensional spinor $\psi$. This single fermion mode can also be understood as the topological zero mode of the Dirac operator $\ii D=\ii D_\vect{x}\cdot\vect{\Gamma}_x+\ii D_\vect{k}\cdot\vect{\Gamma}_k$ in the phase space $T^*M_d$, as required by the index theorem (assuming $M_d=\dsR^d$ is Euclidean):
\eq{
\text{index}(D)=\int_{T^*M_d}\text{ch}(D)=\frac{1}{d!}\int_{\dsR^d\times\dsR^d}\Big(\frac{\dd A}{2\pi}\Big)^d=1.} 
Therefore, regardless of the spatial dimension $d$ of a Fermi sea, its corresponding bulk description as a phase-space Chern insulator is always equivalent to a $(0+1)$-dimensional fermion mode at low energy under dimension reduction. As a result, the classification of the Fermi surface anomaly for a Fermi liquid in $(d+1)$-dimensional spacetime is equivalent to the classification of fermionic SPT phases in $(0+1)$-dimensional spacetime.

The above statement holds true even in the presence of fermion \emph{interactions}. It was originally realized by Teo and Kane \cite{Teo1006.0690} that momentum space (or parameter space) dimensions should be treated as negative dimensions in classifying topological defects in free fermion SPT states. The argument is recently generalized by Jian and Xu \cite{Jian1804.03658} to classify interacting fermionic SPT phases with synthetic dimensions, which is relevant to our discussion here as the Hamiltonian in \eqnref{eq: H synthetic} precisely describes a fermionic SPT system with physical dimension $d$ and synthetic dimension $\delta=d$. According to \refcite{Jian1804.03658}, the key criterion to distinguish the physical and synthetic dimensions relies on the locality of fermion interactions: the interactions must be local in the physical coordinate space and the synthetic momentum space, while non-local in the physical momentum space and the synthetic coordinate space. Based on this principle, the physical momentum is equivalent to the synthetic coordinate in dimension counting; thus, the momentum space dimension should be treated as the synthetic dimension. The main result of \refcite{Jian1804.03658} is that the classification of interacting fermionic SPT states in $(d,\delta)$ physical-synthetic dimension is the same as that in $d_\text{eff}$-dimensional physical space with
\eq{\label{eq: deff}d_\text{eff}=d-\delta.}
Applying this result to our case, we conclude that the classification of interacting phase-space Chern insulators (or phase-space fermionic SPT states more generally) in any spatial dimension $d$ is equivalent to the classification of real-space interacting fermionic SPT states in $(d-\delta)=(d-d)=0$-dimensional space (or, correspondingly, in $(0+1)$-dimensional spacetime).

\subsection{Cobordism Classification Results}

Using the cobordism classification \cite{Kapustin1403.1467, Kapustin1404.6659, Kapustin1406.7329, Freed1604.06527, Guo1711.11587, Wan1812.11967, Yonekura1803.10796, Witten1909.08775, Wan1912.13504, Guo1812.11959} of interacting fermionic SPT states, we propose:
\begin{quote}
The classification of the Fermi surface anomaly associated with the loop group symmetry $\mathrm{L}G$ is equivalent to the classification of $(0+1)$-dimensional interacting fermionic SPT phases with symmetry $G$, which is given by $\text{TP}_1(\Spin\ltimes G)$.
\end{quote}
Here $G$ is the global internal symmetry group, and $\Spin\ltimes G$ denotes the total spacetime-internal symmetry group given by the extension $1\to G\to \Spin\ltimes G\to \Spin\to 1$, with $\Spin$ being the spin group of the spacetime manifold. Although we start with the Dirac fermion theory \eqnref{eq: H Dirac} in the $2d$-dimensional phase space, the effective Euclidean spacetime manifold is only $(0+1)$-dimensional after the synthetic dimension reduction, so the Euclidean spacetime rotation symmetry of the fermionic spinor field is described by the $\Spin(1)$ group. In the presence of time-reversal symmetry, the $\Spin$ structure can be further extended to $\mathrm{Pin}^{\pm}$ structures \cite{Wan1912.13504}. The Fermi surface $\partial\Omega$ with symmetry $G$ can have an emergent loop-$\partial\Omega$ group of $G$ symmetry denoted as $\mathrm{L}G$ in general. The notion of loop group symmetry is more subtle when $G$ is discrete, which will be discussed case by case later.

\begin{table}[htp]
\caption{Cobordism classification of the Fermi surface anomaly of the loop group symmetry $\mathrm{L}G$ by $\text{TP}_1(\Spin\ltimes G)$. In the table, $n\in\dsN$ stands for any natural number, and $\Spin\times_H G:=(\Spin\times G)/H$ denotes the quotient of the group product by their shared normal subgroup. $\dsZ_2^F$ denotes the Fermion parity symmetry.}
\begin{center}
\begin{tabular}{c|cc|c}
\hline\hline
$\mathrm{L}G$ & $G$ & $\Spin\ltimes G$ & $\text{TP}_1$\\ 
\hline
$\mathrm{L}\U(1)$ & $\U(1)$ & $\Spin^c$ & $\dsZ$ \\
$\mathrm{L}\U(n)$ & $\U(n)$ & $\Spin\times_{\dsZ_2^F}\U(n)$ & $\dsZ$\\
$\mathrm{L}\SU(2n)$ & $\SU(2n)$ & $\Spin\times_{\dsZ_2^F}\SU(2n)$ & $0$\\
$\tilde{\mathrm{L}}\U(1)\times\dsZ_{2n}$ & $\dsZ_{2n}$ & $\Spin\times_{\dsZ_2^F}\dsZ_{2n}$ & $\dsZ_{2n}$ \\
$\mathrm{L}\SU(2n+1)$ & $\SU(2n+1)$ & $\Spin\times\SU(2n+1)$ & $\dsZ_2$\\
$\tilde{\mathrm{L}}\U(1)\times\dsZ_{2n+1}$ & $\dsZ_{2n+1}$ & $\Spin\times\dsZ_{2n+1}$ & $\dsZ_{4n+2}$ \\
$\LU(1)\rtimes\dsZ_2^T$ & $\U(1)\rtimes\dsZ_2^T$ & $\mathrm{Pin}^{-}\ltimes_{\dsZ_2^F}\U(1)$ & $\dsZ$ \\
$\LU(1)\rtimes_{\dsZ_2^F}\dsZ_4^{TF}$ & $\U(1)\rtimes_{\dsZ_2^F}\dsZ_4^{TF}$ & $\mathrm{Pin}^{+}\ltimes_{\dsZ_2^F}\U(1)$ & $\dsZ$ \\
$\LU(1)\times\dsZ_2^T$ & $\U(1)\times\dsZ_2^T$ & $\mathrm{Pin}^c$ & 0 \\
\hline\hline
\end{tabular}
\end{center}
\label{tab: class}
\end{table}

In $(0+1)$-dimensional spacetime, SPT phases protected by the total symmetry $\Spin\ltimes G$ are classified by the cobordism group $\text{TP}_1(\Spin\ltimes G)$ \cite{Freed1604.06527} and their topological invariants are given by the cobordism group generators (i.e., the cobordism invariants). Here $\text{TP}$ is shorthand for the topological phase \cite{Freed1604.06527,Guo1711.11587,Wan1812.11967,Wan1912.13504}. \tabref{tab: class} summarizes a few examples of the cobordism classification of Fermi surface anomalies.
The cobordism group element $k\in\text{TP}_1(\Spin\ltimes G)$ is always an integer index given by
\eq{k=\pm q N,}
where $q$ is the symmetry charge carried by the fermion, $N$ is the multiplicity (flavor degeneracy) of the Fermi surface and the sign depends on whether the Fermi surface is electron-like ($+$) or hole-like ($-$). If there are multiple Fermi surfaces in the system, each one can have an independent integer-valued cobordism index $k_\alpha\in\text{TP}_1(\Spin\ltimes G)$. The total Fermi surface anomaly is characterized by a $\U(1)$-valued index $\nu$, 
\eq{\nu=\sum_{\alpha}k_\alpha\frac{\vol\Omega_\alpha}{(2\pi)^d}\mod 1.}
Each cobordism index $k_\alpha$ is multiplied by the fraction of Fermi volume $\vol\Omega_\alpha$ in the Brillouin zone. The Fermi surface anomaly can vanish in the following cases:
\begin{itemize}
\setlength\itemsep{0pt}
\item $\vol\Omega_\alpha/(2\pi)^d\in\dsZ$. The Fermi volume is an integer multiple of the Brillouin zone volume, i.e., the fermion filling is an integer per unit cell for every fermion flavor. In this case, there is no Fermi surface anomaly regardless of the cobordism index $k$.
\item $k_\alpha\sim 0$ (meaning $k_\alpha=0$ when $k\in\dsZ$ or $k_\alpha=0\mod 2n$ when $k\in\dsZ_{2n}$). When the cobordism index $k_\alpha$ is trivial, there is no Fermi surface anomaly, regardless of the filling. This scenario becomes particularly noteworthy when the cobordism group is $\dsZ_{2n}$, as in this case $k_\alpha=2n$ multiples of the (unit-charged) Fermi surface can collectively cancel the anomaly and become deformable to a symmetric product state. 
\item Multiple Fermi surfaces of different cobordism indices $k_\alpha$ and Fermi volumes $\vol\Omega_\alpha$ can cancel the anomaly collectively, if $\nu$ adds up to an integer. Examples of such have been recently studied in the context of Fermi surface symmetric mass generation (SMG) \cite{Lu2210.16304}.
\end{itemize}
More generally, the Fermi surface SMG refers to the phenomenon that the Fermi surface anomaly vanishes $\nu\sim 0$. Still, no symmetric fermion bilinear operator can gap out the Fermi surface into a symmetric product state. Then the symmetric gapping of the Fermi surface can only be achieved through non-trivial interaction effects. It generalizes the concepts of the interaction-reduced SPT classification \cite{Fidkowski0904.2197,Fidkowski1008.4138,Turner1008.4346,Ryu1202.4484,Qi1202.3983,Yao1202.5805,Gu1304.4569,Wang1401.1142,Metlitski1406.3032,You1409.0168,Cheng1501.01313,Yoshida1505.06598,Gu1512.04919,Song1609.07469,Queiroz1601.01596,Witten1605.02391,Wang1703.10937,Kapustin1701.08264,Wang2018Tunneling1801.05416,Wang1811.00536,Gaiotto1712.07950,Aasen2109.10911,Barkeshli2109.11039,Manjunath2210.02452,Zhang2211.09127,Zhang2204.05320} and symmetric mass generation \cite{Wang1307.7480,Ayyar1410.6474,Slagle1409.7401,BenTov1412.0154,Catterall1510.04153,Ayyar1511.09071,Catterall1609.08541,Ayyar1606.06312,Ayyar1611.00280,He1603.08376,DeMarco1706.04648,Ayyar1709.06048,You1705.09313,Schaich1710.08137, Kikukawa1710.11101, Kikukawa1710.11618, You1711.00863,Catterall1708.06715,Butt1811.01015,Butt1810.06117,
Wang1809.11171, Catterall2002.00034, Xu2103.15865, Tong2104.03997, Catterall2010.02290, Butt2101.01026, Butt2111.01001, Zeng2202.12355, Wang2204.14271} to the case of finite fermion filling. We will explore more examples of such in the next subsection.

\subsection{Examples and Comments}\label{sec: cases}

In the following, we will provide some physical understanding of the cobordism classifications in several different cases. To focus our discussion on the \emph{discrete} aspect of the Fermi surface anomaly (as characterized by the integer-valued cobordism index $k$), we will restrict our scope to a unit-charged ($q=1$) single Fermi surface of multiplicity $N$ (such that the cobordism index is $k=N$) with a generic Fermi volume $\vol \Omega$ (e.g.~$\vol \Omega$ is some irrational fraction of the Brillouin zone volume), such that the Fermi surface anomaly index $\nu=k\vol \Omega/(2\pi)^d$ is only trivialized when the cobordism index $k\sim0$ belongs to the trivial class. 

Our starting point will be the dimension-reduced $(0+1)$-dimensional effective bulk theory of the Fermi liquid, as described by the single-mode fermion Hamiltonian \eqnref{eq: H0d}. The objective is to understand the interacting fermionic SPT classification in this $(0+1)$-dimensional quantum system and make connections to the classification of Fermi surface anomaly. After the case by case discussions, we summarize the anomaly-free condition in \tabref{tab: summary}.

\subsubsection{$G=\U(1)$ and $\dsZ$ Classification} 

The $G=\U(1)$ is the most common symmetry in the conventional discussion of Fermi liquids, under which the fermion operator $\psi$ transforms as $\psi\to\e^{\ii\phi}\psi$ for $\phi\in[0,2\pi)$. The dimension-reduced bulk effective Hamiltonian $H=m(\psi^\dagger\psi-1/2)$ has only two eigenstates: $\ket{n_\psi=0}$ and $\ket{n_\psi=1}$, labeled by the two distinct eigenvalues of the fermion number operator $n_\psi:=\psi^\dagger\psi$. The excitation gap closes at $m=0$ as the ground state switches from one to another, which is also the point where the particle-hole symmetry $\dsZ_2^C$ is restored. The gap closing signifies a ``quantum phase transition'' in the $(0+1)$-dimensional system. Therefore, $m<0$ and $m>0$ should be identified as two different SPT phases. If there are many copies of such system, each copy can undergo the SPT transition separately, leading to $\dsZ$-classified SPT phases.

In the presence of the $\U(1)$ symmetry, this gap closing can not be avoided even under interaction. Because the $\U(1)$ symmetry enforces that the interaction can only take the form of a polynomial of $n_\psi$, which does not change the fact that $\ket{n_\psi=0}$ and $\ket{n_\psi=1}$ are still eigenstates of the interacting Hamiltonian. Then the two states have to degenerate on the locus of $\langle n_\psi\rangle=1/2$ where the particle-hole symmetry $\dsZ_2^C$ is restored, resulting in the unavoidable gap closing. So the $\dsZ$ classification is robust against fermion interaction, confirming the cobordism calculation.

As discussed previously in \secref{sec: blk}, the Fermi surface should be defined as the particle-hole symmetric sub-manifold in the phase space. Tuning the mass parameter $m$ across 0 in the effective theory can be viewed as going across the Fermi surface in the momentum space. The inevitable gap closing at $m=0$ (or at the particle-hole symmetric point) corresponds to the protected gapless fermions on the Fermi surface. The cobordism index $k\in\dsZ$ labels the number of gapless fermion modes (assuming fermions are unit-charged under $\U(1)$) both at the SPT transition in the effective theory and on the Fermi surface in the Fermi liquid system. 

In this case, the emergent symmetry on the Fermi surface is $\mathrm{L}_{\partial\Omega}\U(1):\psi_{\vect{k}_F}\to\e^{\ii\phi(\vect{k}_F)}\psi_{\vect{k}_F}$, which is defined for any smooth phase function $\e^{\ii\phi(\vect{k}_F)}$ on the Fermi surface $\partial\Omega$. There is no further constraint on the choice of the function $\phi(\vect{k}_F)$. The loop group symmetry is denoted as $\mathrm{L}\U(1)$ for short in \tabref{tab: class}.

\subsubsection{$G=\U(n)$ and $\dsZ$ Classification}

Apart from carrying $\U(1)$ charge, the fermions may also have internal degrees of freedom. For example, electrons also carry the $\SU(2)$ spin freedom, such that for electronic Fermi liquid in a metal, the total internal symmetry is $\U(1)\times_{\dsZ_2^F}\SU(2)=\U(2)$. More generally, we may consider a $\U(n)$ symmetry, under which an $n$-component fermion field $\psi$ transforms as $\psi_a\to U_{ab}\psi_b$ for $U\in\U(n)$. The classification of Fermi surface anomaly for $G=\U(n)$ is the same as that of $G=\U(1)$, which is $\dsZ$, because the protecting symmetry is only the $\U(1)=\U(n)/\SU(n)$ quotient group. In this case, the emergent symmetry on the Fermi surface is $\mathrm{L}_{\partial\Omega}\U(n):\psi_{\vect{k}_F}\to U(\vect{k}_F)\psi_{\vect{k}_F}$ with $U(\vect{k}_F)\in \U(n)$, denoted as $\LU(n)$ in \tabref{tab: class}.

\subsubsection{$G=\SU(2n)$ and Trivial Classification}

However, once the internal symmetry is reduced from $\U(2n)$ to $\SU(2n)$, the classification collapses, and there is no Fermi surface anomaly for any Fermi volume. From the perspective of the $(0+1)$-dimensional effective theory, a $\SU(2n)$ fundamental fermion $\psi$ (which contains $2n$ flavor components $\psi_a$ for $a=1,2\cdots,2n$) can always be gapped by the following multi-fermion interaction,
\eq{\label{eq: H_int}
H_\text{int}=\prod_{a=1}^{2n}\psi_a +\text{h.c.}.} 
This interaction always stabilizes a unique $\SU(2n)$ singlet ground state. In the presence of this interaction, the $m<0$ and $m>0$ phases can be smoothly tuned to each other without gap closing. Therefore, the $(0+1)$-dimensional interacting fermionic SPT states have only a trivial class under the $\SU(2n)$ symmetry.

The vanishing Fermi surface anomaly implies that the $\SU(2n)$ symmetric Fermi liquid at any filling level (of any Fermi volume) can always be deformed into a gapped product state without breaking the $\SU(2n)$ and translation symmetry. For $n=1$, this gapping term is simply the $s$-wave spin-singlet pairing. For $n>1$, the gapping will be achieved by uniform $\SU(2n)$-singlet multi-fermion condensation. Such multi-fermion condensation can happen independently on each site (or in each unit cell), resulting in a gapped symmetric product state.

\subsubsection{$G=\dsZ_{2n}$ and $\dsZ_{2n}$ Classification}

When reducing the symmetry from $\U(2n)$ to $\SU(2n)$, what essentially happens is that the $\U(1)=\U(2n)/\SU(2n)$ quotient group is broken to its $\dsZ_{2n}$ subgroup, which is also the $\dsZ_{2n}$ center of $\SU(2n)$. In fact, we only need to keep this essential $\dsZ_{2n}$ center symmetry, under which the fermion operator transforms as $\psi\to\e^{\frac{2\pi\ii}{2n}m}\psi$ for $m=0,1,\cdots, 2n$. The multi-fermion condensation interaction \eqnref{eq: H_int} is still the gapping interaction to trivialize the SPT phase (or to gap out the SPT phase transition). However, since the $\dsZ_{2n}$ group has only 1-dimensional representations, the fermionic SPT root state (the generator state) only contains one fermion flavor. Therefore, the trivialization is achieved at $2n$ copies of the root state so that the classification is $\dsZ_{2n}$.

In particular, for $n=1$, a $\dsZ_2$ symmetric Fermi liquid allows the opening of a pairing gap by superconductivity. In this case, the $\dsZ_2$ classification of the Fermi surface anomaly indicates that for generic Fermi volume, the deformation of the Fermi liquid to a symmetric product state is only achievable when there are two fermion flavors (like spin-1/2 electrons) with the cobordism index $k=2\sim 0$, which enables the $s$-wave spin-singlet pairing. One may wonder, even when the fermion flavor number is one (like spinless fermions in condensed matter language) with the cobordism index $k=1$, it is still possible to fully gap the Fermi surface by $p_x+\ii p_y$ pairing in $(2+1)$D, although the Fermi surface anomaly is not canceled for general Fermi volume. However, one should note that the $p_x+\ii p_y$ superconductor is not a trivial gaped state, as it is not deformable to a product state due to its chiral edge mode. The non-vanishing Fermi surface anomaly at $k=1$ enforces the non-trivial invertible topological order in the gapped state. This is related to many discussions about filling-enforced SPT states in the literature \cite{Lu1705.04691, Lu1705.09298, Jiang1907.08596}.

Another case worth discussing is the $n=2$ case, which is the simplest case where Fermi surface SMG \cite{Lu2210.16304} can occur. In this case, the fermions have a $\dsZ_4$ internal symmetry that forbids any pairing gap from opening on the fermion bilinear level. The $\dsZ_4$ classification indicates that every four copies of the Fermi surface (with generic Fermi volume) can be deformed to a gapped product state by interaction. A simple lattice model to demonstrate this phenomenon is described by the following Hamiltonian,
\eq{\label{eq: H Z4}
H=\sum_{a=1}^{4}\sum_{ij}t_{ij}\psi_{ia}^\dagger \psi_{ja}+g \sum_{i}\psi_{i1}\psi_{i2}\psi_{i3}\psi_{i4}+\text{h.c.}.}
There are four fermion modes $\psi_{ia}$ ($a=1,2,3,4$) on each site $i$. The $t_{ij}$ term describes a generic fermion hopping model on the lattice. Without fine-tuning the chemical potential, the fermion system will generally fall in the Fermi liquid phase with a generic Fermi volume. Gapping of the Fermi surface can be achieved by the $\dsZ_4$-symmetric interaction $g$, which drives four-fermion condensation on each site, leading to a gapped symmetric product state in the $g\to\infty$ limit. This gapping mechanism applies to lattice fermions in any spatial dimension. So the $\dsZ_4$-symmetric Fermi liquid is universally $\dsZ_4$ classified in any dimension. 

A key feature of our dimension counting argument is that the classification of the Fermi surface anomaly does not depend on the spacetime dimension. Instead, if we naively considered $(d+1)$D Fermi liquid as a quantum Hall insulator in the $2d$-dimensional phase space, we might mistakenly classify the Fermi surface anomaly by fermionic SPT states in $(2d+1)$-dimensional spacetime. The problem may not be exposed if the symmetry is $\U(1)$ because the classification is always $\dsZ$ and never gets reduced by the interaction effect. So we would not tell any difference. However, once the $\U(1)$ symmetry is broken to its $\dsZ_4$ subgroup, the discrepancy will be manifest. Take $d=2$ for example, the phase space is a 4-dimensional space, and the $\dsZ_4$-symmetric fermionic SPT states in $(4+1)$-dimensional spacetime is $\dsZ_{16}$ classified, which clearly deviates from the $\dsZ_{4}$-classified Fermi surface anomaly predicted by our theory. We know that $\dsZ_4$ should be the correct answer because the lattice model \eqnref{eq: H Z4} explicitly trivialized the Fermi surface in multiples of four (not sixteen). This speaks for the correctness of our dimension counting approach that the momentum space should be treated as negative dimensions, and Fermi liquids in any dimension are topologically equivalent to $(0+1)$-dimensional fermionic SPT states (with boundaries).

Finally, we would like to comment on the emergent loop group symmetry on the Fermi surface when the $\U(1)$ symmetry is broken to $\dsZ_{2n}$. With the multi-fermion condensation term $g$, the low-energy theory takes the form of
\eqs{&H=\sum_{\vect{k}_F\in\partial\Omega}\epsilon_{\vect{k}_F}\psi_{\vect{k}_F}^\dagger\psi_{\vect{k}_F}+\cdots\\&+g\sum_{\{\vect{k}_F^{(a)}\}\in\partial\Omega}\delta_{\sum_{a=1}^{2n}\vect{k}_F^{(a)}}\prod_{a=1}^{2n}\psi_{\vect{k}_F^{(a)}}+\text{h.c.},}
which is symmetric under
\eq{\psi_{\vect{k}_F}\to\e^{\ii\frac{2\pi p}{2n}}\e^{\ii\phi(\vect{k}_F)}\psi_{\vect{k}_F},}
with $p=0,1,\cdots,2n$ labeling a $\dsZ_{2n}$ group element and $\phi(\vect{k}_F)\sim\phi(\vect{k}_F)+2\pi$ being a smooth function of $\vect{k}_F$ subject to the following constraint:
\eq{\label{eq: constraint}\forall \sum_{a=1}^{2n}\vect{k}_F^{(a)}=0:\sum_{a=1}^{2n}\phi(\vect{k}_F^{(a)})=0\mod 2\pi.}
All the $\U(1)$ functions $\e^{\ii\phi(\vect{k}_F)}$ satisfying the constraint in \eqnref{eq: constraint} form a group under pointwise multiplication. We denoted this constrained loop group as $\tilde{\mathrm{L}}_{\partial\Omega}\U(1)$. Then the emergent symmetry on the Fermi surface is $\tilde{\mathrm{L}}_{\partial\Omega}\U(1)\times\dsZ_{2n}$, or short-handed as $\tilde{\mathrm{L}}\U(1)\times\dsZ_{2n}$ in \tabref{tab: class}.

\subsubsection{$G=\SU(2n+1)$ and $\dsZ_2$ Classification}

We have discussed the case of $\SU(2n)$ flavor symmetry with an even number of fermion flavors. Now we turn to the case when the fermion flavor number is odd and the flavor symmetry is $\SU(2n+1)$.  The major difference here is that the $\SU(2n+1)$ flavor symmetry group no longer contains the $\dsZ_2^F$ fermion parity symmetry as a subgroup. In this case, the Fermi surface anomaly is $\dsZ_2$ classified. The physical argument is that with a single copy of the $\SU(2n+1)$ fundamental fermion $\psi$ (with contains $2n+1$ flavor components $\psi_a$ for $a=1,2,\cdots,2n+1$), it is no longer possible to write down the $\SU(2n+1)$-singlet multi-fermion gapping term of the form $\prod_{a=1}^{2n+1}\psi_a+\text{h.c.}$ in the $(0+1)$-dimensional effective theory, because such a term contains an odd number of fermion operators and does not respect the $\dsZ_2^F$ fermion parity symmetry. Therefore, one has to double the system and introduce two $\SU(2n+1)$ fundamental fermions $\psi_1$ and $\psi_2$, such that the following gapping interaction becomes possible
\eq{\label{eq: Hint 2n+1}
H_\text{int}=\prod_{a=1}^{2n}\psi_{1a}\psi_{2a}+\text{h.c.}.}
Similar multi-fermion interaction is applicable to gap out the Fermi surface at a generic Fermi volume if there are two copies of $\SU(2n+1)$ fundamental fermions on the Fermi surface, which explains the $\dsZ_2$ classification. This is also an example of the Fermi surface SMG.

\subsubsection{$G=\dsZ_{2n+1}$ and $\dsZ_{4n+2}$ Classification}

If the $\SU(2n+1)$ flavor symmetry is broken to its center $\dsZ_{2n+1}$ symmetry group, under which the fermion operator transforms as $\psi\to\e^{\frac{2\pi\ii}{2n+1}m}\psi$ for $m=0,1,\cdots, 2n+1$, the Fermi surface anomaly classification will be $\dsZ_{4n+2}$. The physics is essentially the same as the $G=\SU(2n+1)$ case, which relies on the same multi-fermion interaction \eqnref{eq: Hint 2n+1} to drive the SMG in the $(0+1)$-dimensional effective theory. Similar interaction also drives Fermi surface SMG. The SMG gapping mechanism only works when the fermion flavor number is $4n+2$, which is consistent with the $\dsZ_{4n+2}$ classification.

\subsubsection{$G=\U(1)\rtimes\dsZ_2^T$ and $\dsZ$ Classification}

We can extend our discussion to anti-unitary symmetries \cite{Wigner1960a,Wigner1960b}, which will be generally denoted as time-reversal symmetries $\dsZ_2^T$. There are different ways that an anti-unitary symmetry can be combined with the $\U(1)$ charge conservation symmetry of the fermion. Let us first consider the case of $G=\U(1)\rtimes\dsZ_2^T$, where the $\U(1)$ rotation does not commute with the anti-unitary symmetry action $\scT\in\dsZ_2^T$ and $\scT^2=+1$. More specifically, the fermion operator transforms as
\eqs{\U(1)&:\psi\to\e^{\ii\phi}\psi,\\
\dsZ_2^T&:\psi\to\scK\psi, \psi^\dagger\to\scK\psi^\dagger,}  
where $\scK\ii\scK^{-1}=-\ii$ denotes the complex conjugation operator that acts on all complex coefficients in the operator algebra.

In this scenario, the presence of the anti-unitary symmetry does not alter the anomaly classification. The $(0+1)$-dimensional effective theory, characterized by the Hamiltonian $H=m(\psi^\dagger\psi-1/2)$, still includes the mass term $m$ which is symmetric under $\dsZ_2^T$. As the anti-unitary symmetry does not impose additional restrictions on the Hamiltonian, the SPT classification remains unchanged from the case with $G=\U(1)$, which is $\dsZ$. As a result, the Fermi surface anomaly is still classified as $\dsZ$.

\subsubsection{$G=\U(1)\rtimes_{\dsZ_2^F}\dsZ_4^{TF}$ and $\dsZ$ Classification}

Another way to combine the anti-unitary symmetry with $\U(1)$ is to consider $G=\U(1)\rtimes_{\dsZ_2^F}\dsZ_4^{TF}$, meaning that the $\U(1)$ rotation does not commute with the generator $\scT\in \dsZ_4^{TF}$ of the anti-unitary symmetry, but $\scT^2=-1$ (or more precisely, $\scT$ squares to the fermion parity operator, hence the anti-unitary symmetry is four-fold and sharing the $\dsZ_2^F$ subgroup with $\U(1)$). This is actually the standard time-reversal symmetry of electrons that enforces a Kramers doublet \cite{Kramers1930}. The fermion operator $\psi=(\psi_\uparrow, \psi_\downarrow)^\intercal$ is a doublet, which transforms under the symmetry as
\eqs{\U(1)&:\psi\to\e^{\ii\phi}\psi,\\
\dsZ_4^{TF}&:\psi_\uparrow\to\scK\psi_\downarrow, \quad \psi_\downarrow\to\scK\psi_\uparrow.}
The time-reversal symmetry is denoted as a $\dsZ_4$ group because its two-fold action is non-trivial and corresponds to the fermion parity operation ($\psi\to-\psi$) that falls in the $\dsZ_2^F$ subgroup of $\U(1)$.

The mass term $H=m(\psi^\dagger\psi-1/2)$ is still allowed in the effective Hamiltonian under the $\dsZ_4^{TF}$ symmetry. As the anti-unitary symmetry does not introduce new restrictions, the SPT classification remains the same as the $G=\U(1)$ case, which is $\dsZ$. Therefore, the Fermi surface anomaly is also $\dsZ$ classified in this case.

\subsubsection{$G=\U(1)\times\dsZ_2^T$ and Trivial Classification}

We further consider $G=\U(1)\times\dsZ_2^T$ where the $\dsZ_2^T$ anti-unitary symmetry operation commutes with the $\U(1)$ symmetry operation. The symmetry action can be realized on the fermion operator as
\eqs{\U(1)&:\psi\to\e^{\ii\phi}\psi,\\
\dsZ_2^T&:\psi\to\scK\psi^\dagger, \psi^\dagger\to\scK\psi.}
The anti-unitary symmetry $\dsZ_2^T$ here should be interpreted as a particle-hole symmetry, which maps $\psi$ and $\psi^\dagger$ to each other.

In the presence of this symmetry, the original mass term $H=m(\psi^\dagger\psi-1/2)$ is forbidden in the effective Hamiltonian. A symmetry-allowed mass term can only be realized in the doubled system, where the fermion operator $\psi=(\psi_+,\psi_-)^\intercal$ must contain two components, and the symmetry $G=\U(1)\times\dsZ_2^T$ acts as
\eqs{\U(1)&:\psi_\pm\to\e^{\ii\phi}\psi_\pm,\\
\dsZ_2^T&:\psi_\pm\to\scK\psi_\mp^\dagger, \psi_\pm^\dagger\to\scK\psi_\mp,}
such that two anti-commuting mass terms are allowed
\eq{H=m(\psi_{+}^\dagger\psi_{+} - \psi_{-}^\dagger\psi_{-})+m'(\ii\psi_{-}^\dagger \psi_{+}+\text{h.c.}).}
It is possible to tune smoothly from $m<0$ to $m>0$ without closing the excitation gap of this $(0+1)$-dimensional system in the presence of $m'\neq 0$. Therefore, all gapped state belongs to the same SPT phase and the SPT classification is trivial. 

Mapping to the Fermi surface, imposing the particle-hole symmetry enforces the Fermi surface to be perfectly nested \cite{Virosztek1990PRB}. Tuning $m$ from the inside ($m<0$) to the outside ($m>0$) of the Fermi surface, two bands cross at the Fermi level.  In this case, a band hybridization term (similar to $m'$) is sufficient to gap out the Fermi surface fully without symmetry breaking (note that the nesting momentum is already zero in this case). Therefore, the system is free of Fermi surface anomaly, consistent with the trivial classification.

\section{Summary}\label{sec: summary}

In this work, we propose an approach to classify the Fermi surface anomaly by leveraging the correspondence between the Fermi liquid and the Chern insulator in the phase space. Specifically, we suggest using the classification of interacting fermionic symmetry-protected topological (SPT) states in the phase space to determine the Fermi surface anomaly. The non-commutative geometry of the phase space implies that the phase-space SPT states follow unusual dimension counting, where the momentum space dimensions are treated as negative dimensions. As a result, the effective spacetime dimension for the classification problem is reduced to $(0+1)$D. To support our argument, we analyze a phase-space Dirac fermion field theory of fermionic SPT states and apply the dimension reduction technique after resolving the non-commutative geometry. Our proposed approach offers a comprehensive and rigorous way to classify the Fermi surface anomaly, providing valuable insights into the universal low-energy features of electrons in metals.

\begin{table}[hbtp]
\caption{The summary of the Fermi surface anomaly-free condition. The system with a certain number of copies can be symmetrically gapped. The anomaly is free if and only if $\nu = k \frac{\vol\Omega}{(2\pi)^d} =0 \mod 1$. The integer-valued index $k$ is classified by cobordism in \tabref{tab: class}. The case that the normalized Fermi volume $\frac{\vol\Omega}{(2\pi)^d}$ is irrational is discussed in \secref{sec: cases} and summarized in the fourth column. The condition for the normalized Fermi volume being rational number $p/q$ with $p,q\in \dsZ$ is summarized in the third column. }
\begin{center}
\begin{tabular}{c|c|c|c}
\hline \hline
\multirow{2}{*}{$\mathrm{L}G$} & integer & \multicolumn{2}{c}{Number of copies to trivialize} \\ \cline{3-4}
&  index $k$ & $\frac{\vol\Omega}{(2\pi)^d}=p/q$ & $\frac{\vol\Omega}{(2\pi)^d}$ is irrational \\ \hline

$\mathrm{L}\U(1)$ &  $\dsZ$ & $q$ & Never \\
$\mathrm{L}\U(n)$ &  $\dsZ$ & $q$ & Never\\
$\mathrm{L}\SU(2n)$ &  $0$  & $1$ & 1\\
$\tilde{\mathrm{L}}\U(1)\times\dsZ_{2n}$ &  $\dsZ_{2n}$ & $\gcd(q,2n)$ & $2n$ \\
$\mathrm{L}\SU(2n+1)$ &  $\dsZ_2$ & $\gcd(q,2)$ & 2\\
$\tilde{\mathrm{L}}\U(1)\times\dsZ_{2n+1}$ &  $\dsZ_{4n+2}$ & $\gcd(q,4n+2)$ & $4n+2$ \\
$\LU(1)\rtimes\dsZ_2^T$ & $\dsZ$ & $q$ & Never\\
$\LU(1)\rtimes_{\dsZ_2^F}\dsZ_4^{TF}$ &  $\dsZ$ & $q$ & Never \\
$\LU(1)\times\dsZ_2^T$ &  0 & $1$ & $1$\\
\hline \hline

\end{tabular}
\end{center}
\label{tab: summary}
\end{table}

To summarize, the Fermi surface anomaly can be defined by the projective representation of the internal symmetry $G$ on the interstitial defect in the fermion system. It is characterized by a $\U(1)$-valued anomaly index
\eq{\nu=\sum_{\alpha}k_\alpha\frac{\vol\Omega_\alpha}{(2\pi)^d}\mod 1,}
which is a sum of contributions from each Fermi surface labeled by $\alpha$. Each term in the summation contains an integer-valued index $k_\alpha$ multiplied with a real-valued fraction $\vol\Omega_\alpha/(2\pi)^d$. The ratio $\vol\Omega_\alpha/(2\pi)^d$ describes the fraction of Fermi volume $\vol\Omega_\alpha$ in the Brillouin zone. The integer $k_\alpha=\pm q_\alpha N_\alpha$ is given by the fermion charge $q_\alpha$ and multiplicity (flavor degeneracy) $N_\alpha$ of the Fermi surface and classified by the cobordism group $\text{TP}_1(\Spin\ltimes G)$. Assuming a generic Fermi volume for each Fermi surface (i.e.~$\vol\Omega_\alpha/(2\pi)^d$ is not a rational number), the Fermi surface anomaly is determined by the cobordism index $k_\alpha\in \text{TP}_1(\Spin\ltimes G)$. The classification result for a list of internal symmetries $G$ is shown in \tabref{tab: class}.

The complete gapping of the Fermi surface into a product state is feasible if and only if the Fermi surface anomaly vanishes, i.e.~$\nu\sim0$. This can occur through the opening of a superconducting gap (when $G=\dsZ_2$) or a perfect-nested band hybridization gap (when $G=\U(1)\times\dsZ_2^T$) at the free fermion level, when the fermion flavor number falls in the trivial cobordism class. Nevertheless, unconventional mechanisms exist for gapping, referred to as the Fermi surface symmetric mass generation (SMG) \cite{Lu2210.16304}, that can solely be realized via interaction effects when the Fermi surface anomaly vanishes but no fermion bilinear gapping term is allowed due to symmetry constraints. One informative example of such is the quartet (charge-4e) fermion condensation \cite{KivelsonPRB1990, Kameicond-mat/0505468, Berg0810.1564, Radzihovsky0812.3945, Berg0904.1230, Herland1006.3311, Moon1202.5389, Jiang1607.01770, 2021NatPh17.1254G.2021, 2022arXiv220110352G},  on Fermi surfaces with internal $G=\dsZ_4$ symmetry, where the Fermi surface anomaly is $\dsZ_4$ classified. In this scenario, every four multiples of Fermi surfaces can be collectively gapped via four-fermion interactions. The fact that this gapping mechanism is feasible in all dimensions aligns with our assertion that the Fermi surface anomaly is universally categorized by $(0+1)$-dimensional fermionic SPT phases. More cases of Fermi surface trivialization are summarized in \tabref{tab: summary}.

It is worth mentioning that we have only focused on the codimension-1 Fermi surface in this work. However, the synthetic dimension reduction argument in \eqnref{eq: deff} applies to more general codimension-$p$ Fermi surfaces. Assuming the spatial dimension is $d$, a codimension-$p$ Fermi surface will be a $(d-p)$-dimensional closed manifold in the momentum space, which is the boundary of a $\delta=(d-p+1)$-dimensional Fermi sea. As the momentum space (Fermi sea) dimension $\delta$ should be considered as negative dimension, the effective spatial dimension $d_\text{eff}$ for SPT classification is $d_\text{eff}=d-\delta=p-1$, and the corresponding effective spacetime dimension is $d_\text{eff}+1=p$. Therefore, we propose:
\begin{quote}
The codimension-$p$ Fermi surface anomaly with the loop group symmetry $LG$ is classified by $G$-symmetric interacting fermionic SPT phases in $p$-dimensional spacetime, which is given by $\mathrm{TP}_p(\Spin\ltimes G)$.
\end{quote}
For example, consider $\LU(1)$ symmetric generalized Fermi surfaces in $d=3$ dimension, classifications of Fermi surface anomalies are summaried in \tabref{tab: codim}. The results are consistent with the understanding that Fermi rings are topologically unprotected with $\U(1)$ symmetry only, but Weyl points and Fermi surfaces are topologically stable.

\begin{table}[htp]
\caption{Classification of codimension-$p$ Fermi surface anomaly with $\LU(1)$ symmetry in $(3+1)$D spacetime.}
\begin{center}
\begin{tabular}{c|cc|c|cc}
\hline\hline
 & \multicolumn{2}{c|}{Fermi surface} & Fermi sea & & \\
 & codim $p$ & dim & dim & $d_\text{eff}+1$ & $\mathrm{TP}_p$\\
\hline
Weyl points & $3$ & $0$ & $1$ & $2+1$ & $\dsZ\times\dsZ$ \\
Fermi rings & $2$ & $1$ & $2$ & $1+1$ & $0$ \\
Fermi surfaces & $1$ & $2$ & $3$ & $0+1$ & $\dsZ$ \\
\hline\hline
\end{tabular}
\end{center}
\label{tab: codim}
\end{table}%

The classification of Fermi surface anomalies can help us understand the possible ways a Fermi surface can be gapped and the role of interactions in this process. The recent proposal of the ancilla qubit approach \cite{Zhang2001.09159, Zhang2006.01140} for pseudo-gap physics draws a connection between the pseudo-gap metal to Fermi liquid transition with the Fermi surface SMG transition in the ancilla layers, as both transitions are described by field theories of fermionic deconfined quantum critical points \cite{You1705.09313, You1711.00863, Zou2002.02972, Zou2004.14391, Hou2212.13364}. The Fermi surface anomaly constrains the dynamical behavior of such field theories and can potentially shed light on the open problem of pseudo-gap transition in correlated materials.

\begin{acknowledgments}
We acknowledge the discussions with Xiao-Liang Qi, Cenke Xu, Chao-Ming Jian, Chong Wang, Meng Cheng, Nathan Seiberg, Dominic Else, Ryan Thorngren, Zhen Bi, Umang Mehta, Ashvin Vishwanath, Charles Kane, Ya-Hui Zhang, Subir Sachdev, John McGreevy. DCL and YZY are supported by the National Science Foundation (NSF) Grant DMR-2238360 ``Theoretical and Numerical Investigation of Symmetric Mass Generation''. JW is supported by the Center for Mathematical Sciences and Applications at Harvard University and
NSF Grant DMS-1607871 ``Analysis, Geometry and Mathematical Physics.'' 
\end{acknowledgments}

\bibliography{ref}
\end{document}